\documentclass{egpubl}
\usepackage{sca2022}
\usepackage[utf8]{inputenc}
\DeclareUnicodeCharacter{0301}{***************************************************}

\SpecialIssuePaper         

\usepackage[T1]{fontenc}
\usepackage{dfadobe}  

\biberVersion
\BibtexOrBiblatex
\usepackage[backend=biber,bibstyle=EG,citestyle=alphabetic,backref=true]{biblatex} 
\electronicVersion
\PrintedOrElectronic

\ifpdf \usepackage[pdftex]{graphicx} \pdfcompresslevel=9
\else \usepackage[dvips]{graphicx} \fi

\usepackage{egweblnk} 
\usepackage{amsfonts} 
\usepackage{amsmath} 
\usepackage{physics}
\usepackage{tikz}
\usepackage{mathdots}
\usepackage{yhmath}
\usepackage{cancel}
\usepackage{color}
\usepackage{siunitx}
\usepackage{array}
\usepackage{multirow}
\usepackage{amssymb}
\usepackage{gensymb}
\usepackage{tabularx}
\usepackage{extarrows}
\usepackage{booktabs}
\usetikzlibrary{fadings}
\usetikzlibrary{patterns}
\usetikzlibrary{shadows.blur}
\usetikzlibrary{shapes}
\usepackage{algorithm}
\usepackage{algpseudocode}
\floatname{algorithm}{Procedure}

\title[Combining Motion Matching and Orientation Prediction to Animate Avatars] 
      {Combining Motion Matching and Orientation Prediction \\to Animate Avatars for Consumer-Grade VR Devices}
     
\author[J.\,L. Ponton \& H. Yun \& C. Andujar \& N. Pelechano]
{\parbox{\textwidth}{\centering
J.\,L. Ponton \orcid{0000-0001-6576-4528},
H. Yun \orcid{0000-0001-6192-6673},
C. Andujar \orcid{0000-0002-8480-4713} and
N. Pelechano \orcid{0000-0002-1437-245X}
}
\\
{\parbox{\textwidth}{\centering Universitat Politècnica de Catalunya, Barcelona, Spain\\
       }
}
}

\newcommand{\review}[1]{\textcolor{black}{#1}}

\begin{document}

\teaser{
 \includegraphics[width=0.95\linewidth]{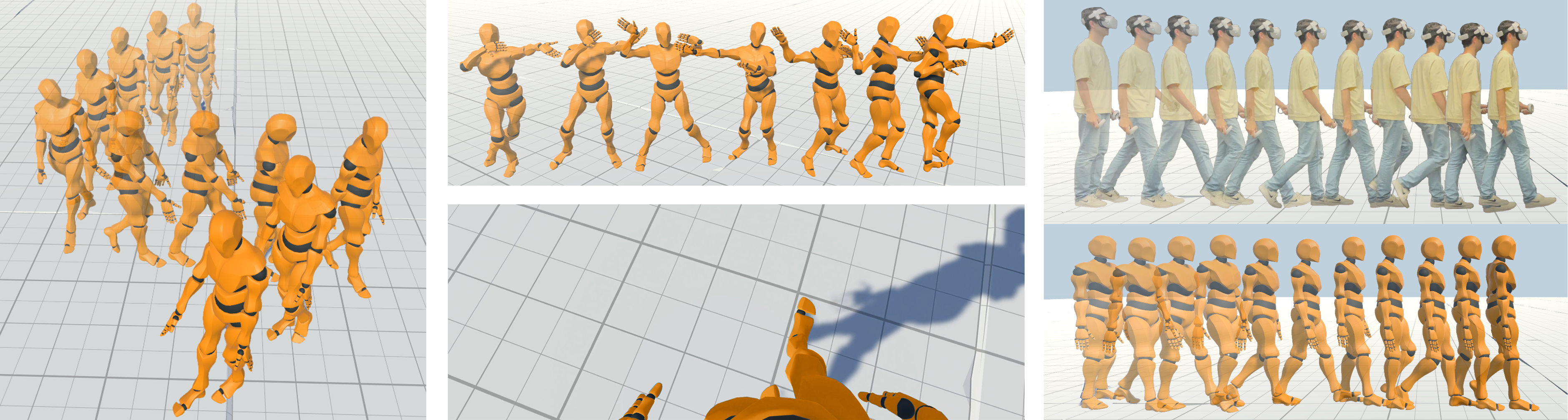}
 \centering
  \caption{An avatar walking and dancing driven by a VR user. The avatar animation is based exclusively on the tracking data provided by the VR headset and two handheld controllers.}
\label{fig:teaser}
}

\maketitle
\begin{abstract}
The animation of user avatars plays a crucial role in conveying their pose, gestures, and relative distances to virtual objects or other users. Self-avatar animation in immersive VR helps improve the user experience and provides a Sense of Embodiment. However, consumer-grade VR devices typically include at most three trackers, one at the Head Mounted Display (HMD), and two at the handheld VR controllers. Since the problem of reconstructing the user pose from such sparse data is ill-defined, especially for the lower body, the approach adopted by most VR games consists of assuming the body orientation matches that of the HMD, and applying animation blending and time-warping from a reduced set of animations. Unfortunately, this approach produces noticeable mismatches between user and avatar movements. In this work we present a new approach to animate user avatars that is suitable for current mainstream VR devices. First, we use a neural network to estimate the user's body orientation based on the tracking information from the HMD and the hand controllers. Then we use this orientation together with the velocity and rotation of the HMD to build a feature vector that feeds a Motion Matching algorithm. We built a MoCap database with animations of VR users wearing a HMD and used it to test our approach on both self-avatars and other users’ avatars. Our results show that our system can provide a large variety of lower body animations while correctly matching the user orientation, which in turn allows us to represent not only forward movements but also stepping in any direction.


\end{abstract}  

\section{Introduction}
The rapid decrease in cost of VR headsets has made this technology available for the general public. Users wearing a HMD should be represented with an animated self-avatar that accurately follows their movements, as this animation is essential to provide presence and a Sense of Embodiment. In order to get correct animations for self-avatars, one option is to rely on high-end motion capture systems such as Xsens, Vicon, or camera-based systems (see \cite{casermanSurvey2020} for a full survey on the topic). However, standard VR devices include at most a HMD and two controllers, thus providing tracking data of the head and hands exclusively. Some games and VR applications simply render floating bodies and hands that follow these trackers, while ignoring the representation and animation of the lower body, for which no tracking data is available. Alternatively, other VR applications do render the lower body, but they simply blend animation cycles applying time warping, or simply drag the avatar without leg movement.  Furthermore, the avatar is typically oriented according to the HMD forward direction. These simplifications lead to incorrect animations that can affect the Sense of Agency (SoA) and thus embodiment, and that convey wrong information about the user’s poses, gestures and placement within the virtual environment. 

In this work we present a system to generate a plausible animation for VR avatars, which is suitable both for self-avatars (seen from a first-person point of view) and collaborators’ avatars (seen from a third-person point of view), and which only requires the tracking data already provided by consumer-grade VR headsets. In the first stage, we use a neural network to estimate the body orientation from the HMD and hand controllers. Then this information, together with the HMD velocity and orientation, is used to build a feature vector to feed a Motion Matching algorithm that searches the best animation in a database. Since user behavior in VR is somewhat specific (e.g., high frequency of looking-around gestures), we generated a specific MoCap database representing users while doing typical walking and turning actions in VR. 
Our system outputs natural-looking lower body animations that can correctly represent the movements of the user while only needing the three standard tracking devices that are included in most VR systems. 

The main contributions of this paper are:
\begin{itemize}
    \item Body Orientation Prediction using a lightweight neural network based on body-size independent features, such as orientation, velocities and angular velocities of the HMD and hand controllers.
    \item Customized Motion Matching for VR so that avatars' lower body movements are adapted according to the velocity and trajectory of the user instead of using a fixed walking/running animation. 
    \item The integration with an IK-based solution for the upper body that results in a seamless animation of the whole body of the user avatar. 
\end{itemize}

Our system can be used to improve the animation of self-avatars and the avatars of other users also wearing a VR headset, providing natural animations that can be used in VR games and collaborative VR to better trust the users' positions and movement in the virtual environment \cite{rios2018users}.


\section{Related Work}

\subsection{Self-avatar control in VR}

The importance of self-avatars has long been recognized from various perspectives, including user performance, distance perception, cognitive load, Sense of Embodiment (SoE) and presence.  
%
With the popularity of HMD-based VR devices, the impact of the avatar’s visual fidelity has drawn much attention. Due to the limited tracking information provided by consumer-grade VR, the floating hands representation has become the most common form of self-avatars in VR applications and games. However, recent studies have shown that hands-only representations provide little SoE. Jung and Hughes \cite{JungandHughes2016} conducted a study with hand-focused tasks to investigate the effect of inferred body parts on the Sense of Body Ownership (SoBO), one of the subcomponents of SoE. Their results suggest that the inferred lower body leads to a higher level of SoBO than no-lower body condition. 
Fribourg et al. \cite{fribourg2020} investigated the effect of the appearance, control and point-of-view of self-avatars on SoE. They found that most users were not satisfied with abstract self-avatars (e.g., five spheres for body extremities) when performing tasks like yoga, walking and kicking.
Galvan et al. \cite{galvandebarba2020} explored the effect of different levels of body parts' animation on Plausibility Illusion and Sense of Control (SoC). They concluded that adding foot tracking to full-body self-avatars increased the SoC of the users the most.  

Additionally, full-body self-avatars and hand-only avatars lead to different behaviors and cognitive loads in VR. Pan and Steed \cite{pan2019a} demonstrated that a full-body avatar could reduce users' cognitive load when performing spatial reasoning tasks, in contrast to hand-only avatars.  Ogawa et al. \cite{ogawa2020} suggested that realistic full-body self-avatars discouraged people more effectively from walking through virtual walls than hand-only representations.

Many researchers have studied the impact of the lower body’s animation fidelity. Some studies indicate that synchronized leg animation based on feet tracking increases presence and SoC \cite{fribourg2020, galvandebarba2020, park2019}, whereas other studies have not found significant differences for leg animations inferred with and without foot tracking \cite{goncalves2022a}. 
Lee et al. \cite{lee2020} also reached a similar conclusion that a prerecorded animation for the lower body and a synchronized animation predicted by a neural network had a similar effect on presence when observed indirectly during walking in place (WIP). However, synchronized leg animation provided higher presence and Body Ownership when users looked down at their lower body during WIP. 
Galvan et al.\cite{galvandebarba2020} observed that feet tracking induced higher SoC compared to procedurally simulated locomotion. Users reported as \emph{disturbing} the fact that for the simulated condition the feet were always aligned with the HMD. For this reason, instead of adding feet tracking we propose a full-body self-avatar animation algorithm combining body direction prediction and Motion Matching to improve the animation fidelity. 

\subsection{Data-driven Animation Control}

Controlling a self-avatar in VR is equivalent to using sparse high-level input to reconstruct the human pose and motion with real-time constraints. Ellis et al. \cite{Ellis2004} demonstrated that less than 16\,ms end-to-end latency is necessary to achieve perceptual stability in virtual environments. On top of the highly under-constrained nature of pose reconstruction, achieving low latency makes self-avatar control in VR a challenging problem. 

Various user input data from consumer-grade devices can be used to synthesize full-body animation for characters in real time. For example, some studies used optical data from egocentric cameras mounted in a baseball cap \cite{xu2019mo}, a HMD \cite{tome2020,yang2022}, controllers \cite{Ahuja2022}, or glasses \cite{zhao2021} to estimate the body pose. 
Egocentric cameras suffer from extreme perspective distortion and self-occlusion that lead to inadequate tracking information for the lower body. Recent studies show that a sparse set of Inertial Measurement Units (IMUs) could accurately reconstruct full-body human motion with an accuracy similar to that of commercial IMU suits. For instance, DIP \cite{huang2018} and TransPose \cite{yi2021} use learning-based methods to accurately reconstruct the full-body pose with 6 IMUs mounted on users’ wrists, knees, head and pelvis. However, IMUs can be impacted by environmental noise, electromagnetic waves and temperature changes \cite{pei2021}. Furthermore, the latency of state-of-the-art solutions based on optical data or IMUs remains high for VR applications.

Several studies propose real-time methods for reconstructing the body pose from consumer-grade VR devices, including HMD, controllers, and trackers. The number of tracked points varies from three to six. In a six-point tracking setting, the head, pelvis, hands and feet are tracked. IK solvers can calculate the rotations of untracked joints and reconstruct the full-body pose provided that enough information about end-effectors is available \cite{FinalIK, Ponton2022}. A more detailed discussion about IK methods for reconstructing the human body in VR can be found in the survey \cite{casermanSurvey2020}. 
Four-point tracking commonly includes HMD, controllers, and an additional tracker on the pelvis \cite{yang2021} or ankles \cite{caserman2019}. This configuration eliminates the infrared occlusion and foot-floor impact problems with feet trackers.  Yang et al. \cite{yang2021} propose a velocity-based recurrent neural network (RNN) model that accurately predicts low-body pose in real time and could generalize to different body shapes. However it still needs an additional tracker on the pelvis and post-processing to eliminate foot sliding. Their 45 fps frame rate is not enough for real-time VR applications.

Three-point tracking, i.e. only HMD and controllers, has been studied to obtain full-body poses in VR. Learning-based methods like variational autoencoders and RNN models \cite{dittadi2021, lin2019} can generate full-body poses from the three-point tracking data. However, while these methods replicate accurate motions for the upper body, but not for the lower body because of the lack of training data with various leg movements or lack of tracking information for the feet. In our work, the locomotion data also covers walking, squatting, and running. Our goal is to provide realistic lower-body motion instead of replicating user leg movements.

CoolMove \cite{ahuja2021} uses $k$-Nearest-Neighbors ($k$-NN) to generate full-body animations in VR, including boxing, basketball, climbing, running and swimming. They extract features from both the motion database and the live input, along with position and orientation of HMD and controllers. Matched motion candidates returned by $k$-NN are blended with proportional weights to form the output pose. Their result demonstrate that the generated poses could lead to higher SoA when observed from a third-person view, but lower embodiment than the IK solution due to the positional error between users’ input and generated poses. Our work avoids this mismatch problem by generating the upper-body and lower-body pose using the IK solver and Motion Matching respectively. Thus the positional error between the user’s and avatar’s hands is reduced.

Some other works investigate the more extreme setting, i.e. using only HMD data \cite{caserman2016, lee2019}. Recent studies use Deep Neural Networks (DNN) to predict the status of leg movement of WIP with the HMD tracking data and then blend leg animations based on the predicted status \cite{lee2020, hanson2019}. Our work can handle real-walking, which is proven better than WIP in terms of motion sickness and SoE \cite{Slater1995}.


\section{Overview}
\begin{figure*}[tbp]
  \centering
  \mbox{} \hfill
  \includegraphics[width=1.0\linewidth]{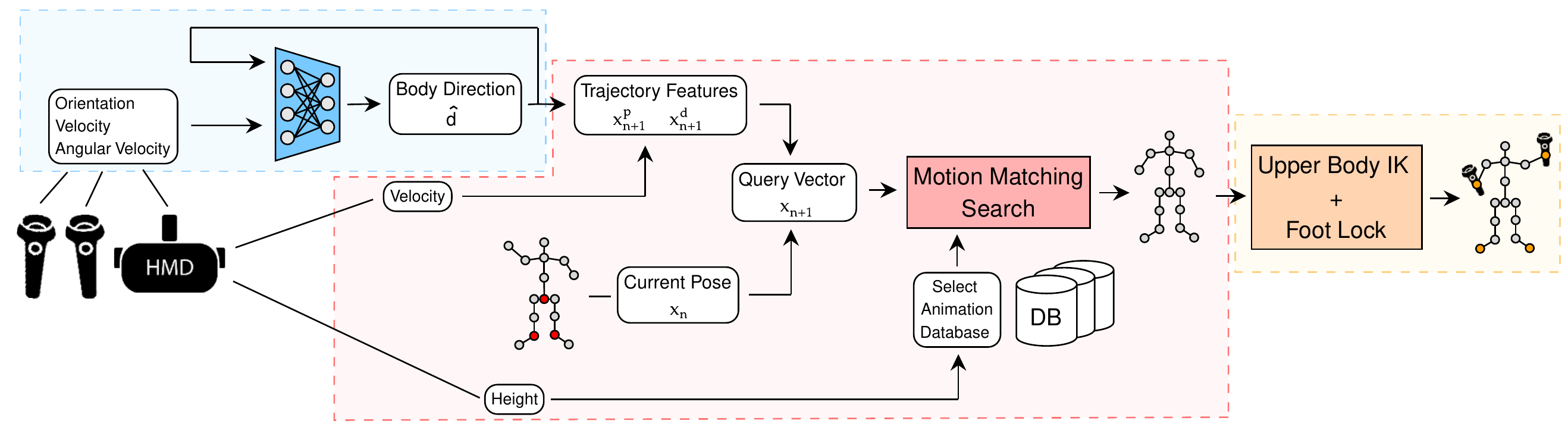}
  \caption{\label{fig:main_pipeline}Our pipeline starts by predicting the body orientation from the rotation, velocity and angular velocity of the HMD and controllers, using a lightweight neural network. In the second step, a query vector is formed by combining predicted trajectory features and pose features extracted from the current pose. The previously predicted orientation is part of the trajectory features. Every 10 frames we use a Motion Matching approach to find the motion that best matches the current pose and trajectory. Finally, we apply IK to adjust the upper body pose to the hand-held controllers.}
\end{figure*}

We propose a new method to realistically animate full-body self-avatars using only the devices included in a typical consumer-grade VR system: one HMD and two hand-held controllers. Figure \ref{fig:main_pipeline} shows an overview of the approach. The core component of our method is Motion Matching, which provides high-quality locomotion animations with seamless transitions between different body orientations and movements (such as idle, walking and running). 

The method can be divided into three parts: body orientation prediction, Motion Matching and final pose adjustments. One of the inputs of Motion Matching is the user's trajectory, which is defined in terms of the future positions and the target body orientation. While the future positions can be predicted from the HMD's velocity, estimating the body orientation from the HMD and controllers is not straightforward. One simple solution would be to use the forward direction of the HMD as the body direction. In that case, Motion Matching will still produce high-quality animations, but the virtual character will continuously change its orientation every time the user rotates the head, as shown in Figure \ref{fig:hmdforward_vs_network}, even when the body remains in the same orientation. Therefore we designed and trained a neural network to obtain better estimates of the actual body orientation, taking as input the rotation, velocity, and angular velocity of all three devices, as well as the previous body orientation. 

Once we have predicted the trajectory of the user, we build a query vector and use Motion Matching to search for the sequence of poses that better match our current pose and target trajectory. Our system supports animations such as crouching by letting the Motion Matching algorithm search in different animation databases, depending on the height of the HMD. Each of these databases have been captured with the performers moving with different levels of leg bend, and the specific database to use is selected at runtime. 

Our method focuses on the animation of the lower body, for which tracking information is missing. In contrast, the arms can be animated by applying IK, since the end effectors can be located and oriented precisely according to the handheld controllers.  
This separation of the upper and lower body parts dramatically decreases the dimensionality of the feature query for the Motion Matching search, while providing a solution that guarantees the correct positioning of the hands at all times. This offers a good trade-off between animation smoothness and pose accuracy.
Motion Matching guarantees smooth transitions as long as we allow the returned avatar positions to drift a little from the requested user trajectory. These drifts are usually imperceptible when controlling the avatar with non-VR devices such as joysticks, but large drifts are clearly perceptible in VR. Therefore we allow users to limit the maximum distance between their actual position and that of their avatars. However, this constraint may introduce some foot sliding, as the root displacement will no longer agree with the chosen animation. We thus apply a foot lock technique to reduce foot sliding.
The following sections describe each part of the method in detail. 


\section{Prediction of the Body Orientation}
\label{sec:directionPrediction}
Predicting the body orientation is a common problem in applications using full-body avatars with only one HMD and two controllers. Often, the forward direction of the HMD is used to orient the whole virtual body. However, this often leads to a significant misalignment between the forward direction (and pose) of the user and the avatar. For example, users may be moving their head to look around (keeping the rest of the body static), but their avatars will rotate the whole body (not just the head) to match the orientation of the HMD.

From the HMD and controllers, we cannot directly extract the orientation of the body. However, we can try to infer it from the information given by these devices. For this purpose, we trained a lightweight feedforward neural network to predict the body orientation from the rotation, velocity and angular velocity of all three devices. The positions are not used so that the method is body-size independent and can be applied to a broader range of users. Our method also takes as an input the previously predicted body orientation to induce temporal continuity.

\subsection{Network Input and Output}
The input vector for the neural network is constructed using the velocities and rotations of the $k=3$ trackers and the previous predicted orientation. It can be defined as $\mathbf{x} = \{ \mathbf{x^v}, \mathbf{x^w}, \mathbf{x^r}, \mathbf{x^d} \} \in \mathbb{R}^{k\times12+6}$ where $\mathbf{x^v} \in \mathbb{R}^{k\times3}$ are the 3D velocities, $\mathbf{x^w} \in \mathbb{R}^{k\times3}$ are the 3D angular velocities (axis-angle rotation vector with the angle encoded as the length of the axis vector), $\mathbf{x^r} \in \mathbb{R}^{k\times6}$ are the 3D rotations, and $\mathbf{x^d} \in \mathbb{R}^{6}$ is the previous predicted orientation. Rotations are represented with the 2-axis rotation matrix used in \cite{zhou2019} to ensure rotation continuity during training. We also normalize each feature of the input by subtracting their mean and dividing by the standard deviation. The output $\mathbf{\hat{d}} \in \mathbb{R}^{6}$ is the predicted body orientation. 

To create the database we used an Xsens motion capture system to record the full body motion of a user while playing different games for SteamVR-based HMDs and Oculus Quest. We also explicitly captured more extreme movements to cover a wide range of poses. In total we used around half a million poses for the training with around 2.4 hours of motion capture. We use this data to simulate the trackers' information. We assume fixed offsets between the head, left and right wrist joints and the corresponding trackers. Then, for each pose, we compute their velocities, angular velocities and rotations. We represent the orientation of the body with the orientation of the virtual root joint computed in Section \ref{sec:pose_database}. To facilitate training, we represent all features with respect to a local coordinate system defined by the following three axes: the projection of the forward direction of the HMD onto the floor plane, the vertical world vector, and their cross product. This guarantees that the prediction is independent of the user’s orientation. 

\subsection{Network Architecture and Training} \label{sec:network_training}
We used a simple feedforward neural network with 2 hidden layers with 32 units each and ReLU activation functions. 
The training is performed as usual with feedforward neural networks. However, predicting the orientation directly from the ground truth data would not match the real usage scenario of the network, and therefore, the network would not be learning how to predict the next orientation based on the previously predicted one. Instead, for every element in a training batch, we iteratively predict the orientation $r$ times (e.g., $r=50$). Then, we compute the MSE loss by comparing the final predicted body orientation $\mathbf{\hat{d}}$ with the ground truth orientation $\mathbf{d^*}$ after $r$ frames. Algorithm \ref{alg:training_body_direction} provides more details about the training process. Notice that subindices in Algorithm \ref{alg:training_body_direction} refer to frames within the motion database used for training.  
The neural network was implemented using PyTorch \cite{pytorch2019} and the hyperparameters tuned using the ASHA scheduler implemented in Ray \cite{liaw2018tune}. We optimized the model with Adam and the final tuned model used a batch size of 64, learning rate of $3 \cdot 10^{-4}$ and weight decay of 0.035.

{
\floatname{algorithm}{Algorithm}
\begin{algorithm}
\caption{Body orientation predictor training}\label{alg:training_body_direction}
\begin{algorithmic}[1]
\Require $D$: body orientations database (training set)
\State $W \gets$ initialize weights
\State $r \gets$ number of iterations (e.g., 50)

\ForAll {epochs}
\State // Iterate $D$ in batches (here batch size = 1 for simplicity)
\For{$i \gets 1$ \textbf{to} $D.length - r - 1$}
    \State // Subindices in this algorithm indicate
    \State // the frame within the database
    \State // Get Previous Body Orientation
    \State $ {\mathbf{x}_{i-1}^\mathbf{d}} \gets$ sample body orientation from $D_{i-1}$
    \State $\mathbf{\hat{d}} \gets {\mathbf{x}_{i-1}^\mathbf{d}}$
    \State // Get Ground Truth Body Orientation
    \State ${\mathbf{x}_{i+r-1}^\mathbf{d}} \gets$ sample body orientation from $D_{i+r-1}$
    \State $\mathbf{d^*} \gets {\mathbf{x}_{i+r-1}^\mathbf{d}}$
    \For {$j \gets 0$ \textbf{to} $r - 1$}
        \State $\mathbf{x^v} \gets$ sample velocity from $D_{i+j}$
        \State $\mathbf{x^w} \gets$ sample angular velocity from $D_{i+j}$
        \State $\mathbf{x^r} \gets$ sample orientation from $D_{i+j}$
        \State // Predict Body Orientation
        \State $\mathbf{\hat{d}} \gets$ Predict$(W, \{ \mathbf{x^v}, \mathbf{x^w}, \mathbf{x^r}, \mathbf{\hat{d}} \})$
    \EndFor
    \State $\mathcal{L} \gets MSE(\mathbf{d^*}, \mathbf{\hat{d}})$
    \State $W \gets$ backpropagate $\mathcal{L}$    
\EndFor
\EndFor
\end{algorithmic}
\end{algorithm}
}

\section{Motion Matching for VR}

Motion Matching is a data-driven algorithm used to animate virtual characters using high-quality motion capture data and a minimal manual setup. 
It was initially presented by B{\"u}ttner and Clavet \cite{buttner2015} and further developed for Ubisoft's game \emph{For Honor} \cite{clavet2016}. Recently, Holden et al. \cite{holden2020} have provided a state-of-the-art Motion Matching implementation used in AAA game productions which we use as the main reference for our implementation. This section presents the details of our adaptation for its use with Head-Mounted Displays in VR.

\subsection{Pose database} \label{sec:pose_database}
Motion Matching searches over an animation database for the best match for the current avatar pose and the predicted trajectory. Since Motion Matching does not create new poses, the animation database is an essential component that determines the quality of the final animations. Again, we used the Xsens motion capture system to capture a wide range of locomotion sequences typically found when a user performs real walking in VR. \review{This includes walking and running forward, backward, different turning rates, in-place rotations, and side stepping.} Compared to other applications, VR requires us to capture slow movements (users tend to walk carefully in VR), with different ranges of velocities for the same movement, and sudden changes in velocity direction and torso orientation. Since the avatar is driven by real users, we cannot enforce the character’s velocity as we could do when using a joystick for a video game, so it is crucial to capture animations at different velocities to ensure that Motion Matching has enough flexibility to follow the user’s trajectory. \review{In total, our animation database contains around 25 thousand poses (approximately 5 minutes of raw MoCap data).}

The animation database can include one or more motion capture files. These files are preprocessed to create the pose database, which is essentially a vector containing all poses in the same order as they appear in the motion capture files but with additional information. Each pose $\mathbf{y}$ is defined as follows:
\begin{equation}
    \mathbf{y} = \left( \mathbf{y^p}, \mathbf{y^r}, \mathbf{y^v}, \mathbf{y^w}, \mathbf{y^c} \right)
\end{equation}
where $\mathbf{y^p}$ are the local joint positions, $\mathbf{y^r}$ are the local joint rotations, $\mathbf{y^v}$ are the local joint velocities, $\mathbf{y^w}$ are the local joint angular velocities and $\mathbf{y^c}$ contains two Boolean values indicating whether the left and right foot are in contact with the ground.

The avatar returned by Xsens has the hip joint as root of the skeleton. Therefore, when creating the pose database, we add a virtual root joint to the skeleton that indicates the position and orientation of the character. It is created by projecting the hip joint onto the floor plane, and its coordinate frame is defined by the projected hip forward direction, the vertical world vector and their cross product. Then the hip joint is transformed to the virtual root space.

\subsection{Feature database}
Notice that $\mathbf{y}$ contains data relative to a given frame of the animation. Since we wish to search for the best sequence of poses, we need to add temporal information. Therefore, instead of directly using the pose database when searching for a new sequence of poses, we compute a new database with the main features defining locomotion \cite{clavet2016}. 
We compute a feature vector $\mathbf{z} \in \mathbb{R}^{27}$ for each pose $\mathbf{y}$. This feature vector combines two types of information: the current pose and the trajectory. When comparing feature vectors, the former ensures no significant changes in the pose and thus smooth transitions; the latter drives the animation towards our target trajectory. Feature vectors are defined as follows:
\begin{equation}
    \mathbf{z} = \left( \mathbf{z^v}, \mathbf{z^l}, \mathbf{z^p}, \mathbf{z^d} \right)
\label{eq:z}
\end{equation}
where $ \mathbf{z^v}, \mathbf{z^l} $ are the current pose features and $ \mathbf{z^p}, \mathbf{z^d} $ are the trajectory features. More precisely, $\mathbf{z^v} \in \mathbb{R}^{9}$ are the velocities of the feet and hip joints, $\mathbf{z^l} \in \mathbb{R}^{6}$ are the positions of the feet joints, $\mathbf{z^p} \in \mathbb{R}^{6}$ and $\mathbf{z^d} \in \mathbb{R}^{6}$ are the future 2D positions and 2D orientations of the character $0.33$, $0.66$ and $1.00$ seconds ahead. Trajectory features are projected onto the ground and all features are local to the virtual root joint, i.e., in character space. Each feature is also normalized by subtracting its mean and dividing by the standard deviation.

\subsection{Search} \label{sec:search}
At runtime, we perform a Motion Matching search every a few frames (e.g., 10 frames). At a given update $n$, the character is at a certain pose described by the feature vector $\mathbf{z}_n$ (including all the elements in Eq.~\ref{eq:z}), and we want to search for the sequence of poses that best matches the predicted user’s trajectory. To do so, we create a query feature vector $\mathbf{q}$ defined as in Eq.~\ref{eq:z} and update it before every search (see Eq.~{\ref{eq:query_vector_1}-\ref{eq:query_vector_4}}). The query vector $\mathbf{q}$ is used to search in the feature database for the closest vector. The returned pose is not necessarily the continuation of the current pose in the animation database. Therefore, although we added temporal information to the feature vectors to prime smooth transitions, there may still be some noticeable changes between poses, especially for motions not included in the database. Therefore, we apply inertialization blending \cite{bollo2017} to smooth the transitions. \review{Since linearly searching over the animation database may be costly, we accelerate the search with a two-layer Bounding Volume Hierarchy as introduced by Holden et al. \cite{holden2020}.}


The trajectory components $\mathbf{q}^\mathbf{p}$, $\mathbf{q}^\mathbf{d}$ are estimated as follows. 
The future position $\mathbf{q}^\mathbf{p}$ is predicted from the HMD’s velocity projected on the ground plane, and the future direction $\mathbf{q}^\mathbf{d}$ is predicted from the trackers’ input data using the neural network described in Section \ref{sec:directionPrediction}.

The HMD velocity $\mathbf{\hat{v}}$ is needed to compute the future positions of the avatar. However, directly using it could lead to undesirable noise and discontinuities. Instead, the velocity should change smoothly so that the search can find suitable trajectory matches. For this purpose, we could use an exponential decay function. We define the smoothed velocity $\mathbf{v}$ as follows:
\begin{equation}
    \mathbf{v}_{n+1} = \mathbf{v}_n + \beta (\mathbf{\hat{v}}_n - \mathbf{v}_n) \Delta t
\end{equation}
where $\Delta t$ is the time between updates and $\beta$ is the responsiveness factor which adjusts how fast $\mathbf{v}$ converges to $\mathbf{\hat{v}}$. 

Similarly, we do not directly use the predicted body orientation $\mathbf{\hat{d}}$ (see Section \ref{sec:directionPrediction}), but a smoothed version $\mathbf{d}$. Although orientations are represented as 2-axis rotation matrices for network training, they can be easily smoothed by representing them as quaternions and using the spherical linear interpolation:
\begin{equation} \label{eq:slerp}
    \mathbf{d}_{n+1} = \mathrm{Slerp}(\mathbf{d_{n}}, \mathbf{\hat{d_{n}}}, \beta \Delta t)
\end{equation}
For simplicity, we show the exponential decay function for velocities and the Slerp function for orientations. However, to obtain a smoother behavior, the results of this paper used a spring-damper-based system \cite{gems4}.

Formally, to create a query vector $\mathbf{q}$ we retrieve the current feature vector $\mathbf{z}_n$ from the feature database, and predict the trajectory of the user from $\mathbf{v}$ and the orientation $\mathbf{d}$:
\begin{align}
    \mathbf{q}^\mathbf{v} &= \mathbf{z}_{n}^\mathbf{v} \label{eq:query_vector_1} \\
    \mathbf{q}^\mathbf{l} &= \mathbf{z}_{n}^\mathbf{l} \label{eq:query_vector_2}\\
    \mathbf{q}^\mathbf{p} &= ( \mathbf{\hat{p}}_n + \frac{1}{3} \mathbf{v}_n, \ \mathbf{\hat{p}}_n + \frac{2}{3} \mathbf{v}_n, \ \mathbf{\hat{p}}_n + \mathbf{v}_n ) \label{eq:query_vector_3}\\
    \mathbf{q}^\mathbf{d} &= ( \mathbf{d}_{n+20}, \ \mathbf{d}_{n+40}, \ \mathbf{d}_{n+60} ) \label{eq:query_vector_4}
\end{align}
where $\mathbf{q}^\mathbf{p}$ and $\mathbf{q}^\mathbf{d}$ contain the future predictions at $0.33$, $0.66$ and $1.00$ seconds assuming the application runs at 60 frames per second. The orientation $\mathbf{d}$ is estimated with Eq.~\ref{eq:slerp} by fixing $\mathbf{\hat{d_{n}}}$ as the current frame prediction. And the target position of the body $\mathbf{\hat{p}}$ is computed by projecting the center of the head on the ground floor (the center is an estimate of the most stable point under rotations of the head). All values are local to the virtual root joint.


\subsection{Position accuracy} \label{sec:position_accuracy}
One of the limitations of Motion Matching is the drift between the desired and actual position and direction of the character. The search tries to find a sequence of poses that follows the target trajectory while avoiding significant changes in the pose. A sequence of poses may better match our target trajectory, but another may be chosen to prevent considerable pose changes. Providing weights for the different features in the query vector $\mathbf{q}$ helps to adjust this quality vs. responsiveness trade-off. However, even if we set to zero the weights for the current pose features, it is impossible to always find a perfect match with the target trajectory since we are constrained by the poses available in the animation database. This is typically not a problem when animating a character in a video game from a third person view, but it can be problematic when animating a self-avatar in VR, where a correct alignment between the virtual character and the user is needed at all times.

We reduce this problem first by slightly moving the virtual root joint towards the target position proportionally to the character's velocity, thus minimizing the adjustment when the character is moving slowly to avoid users noticing the foot sliding introduced. Second, we let users provide a position accuracy parameter $\alpha$ to ensure that the position of the virtual root joint $\mathbf{p}$ does not deviate more than $\alpha$ with respect to the target position $\mathbf{\hat{p}}$ defined in Section \ref{sec:search}. The corrected position of the virtual root joint $\mathbf{p'}$ is computed as follows:
\[
\mathbf{p'}=
\begin{cases}
    \mathbf{\hat{p}} + \alpha \frac{\mathbf{p} - \mathbf{\hat{p}}}{\lVert \mathbf{p} - \mathbf{\hat{p}} \rVert}, & \lVert \mathbf{p} - \mathbf{\hat{p}} \rVert > \alpha \\
    \mathbf{p}, & \text{otherwise}
\end{cases}
\]

If positional accuracy is a priority, users can use a low $\alpha$ (e.g., 10\,cm). This will reduce the positional misalignment that in the case of a self-avatar in VR could lead to a reduction in the Sense of Embodiment. 
On the contrary, a larger value will leave more freedom for Motion Matching to provide higher quality motions, at the expense of some positional drift, which may not be important when animating other users' virtual avatars for collaborative VR. Therefore, $\alpha$ can be adjusted depending on the requirements of the application and the user preference.  

\subsection{Improving motion search for non-upright motions}

One crucial aspect of keeping users immersed in VR is synchronizing the leg movements of the avatar with that of the users. As shown in \cite{Ponton2022}, having the bending of the virtual legs synchronized with the user's legs positively affects the Sense of Embodiment in VR.  Consequently, we provide a way to control the height of the virtual avatar while maintaining fast Motion Matching searches and motion continuity. 

In addition to the standard locomotion database, we captured multiple locomotion databases with different levels of knee bend: from a small bend to having the legs completely bent, or walking on tip-toes. Then, the user's height is represented with a normalized value computed as the ratio between the current HMD's height and the one calculated during a calibration step (at the beginning of the execution, the user is asked to press a button while standing up). Each database has an assigned range of height ratios, and thus, in real-time, we can select the proper database only by querying the HMD's height. 

\begin{figure}[h]
  \centering
  \includegraphics[width=1.0\linewidth]{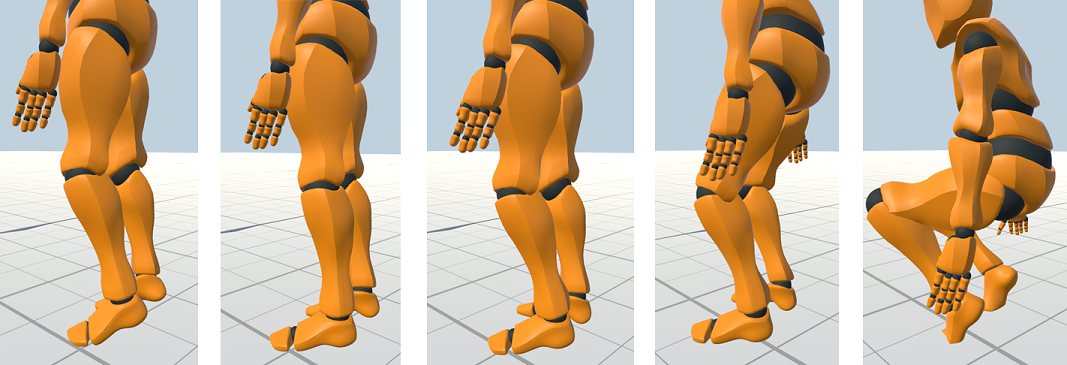}
  \caption{\label{fig:leg_bending} Examples of poses with different leg bends based on the HMD height. From left to right: tip-toes, normal length, small knee bend, medium knee bend, and crouching pose.}
\end{figure}

Every time we change the database, we trigger a Motion Matching search. The query vector $\mathbf{q}$ is computed as usual, and the final result is inertialized to blend significant changes in pose. Although the database changes for the search, by maintaining the current pose features in the query vector (e.g., the local position of the feet), we will obtain a similar pose; for instance, if the right foot is ahead of the left one, the new search will try to find a pose with the same feet configuration in the new database. The result for different databases can be seen in Figure \ref{fig:leg_bending}.

We decided to use different databases to avoid adding more complexity to the search and prevent unnecessary pose changes. While it may be feasible to include the HMD's height in the feature vector, this would have two main drawbacks: First, having an extensive database with all possible motions increases the memory and computation requirements of the search. Second, VR users have complete freedom to move as they wish, which makes the Motion Matching search more challenging. Adding the HMD's height as a feature would drastically hinder finding a suitable pose, and lead to more frequent animation changes.


\section{Final pose adjustments}
In our work, the upper body is not considered for the Motion Matching algorithm to avoid increasing the dimensionality of the feature vector and focus instead on lower body locomotion, for which no tracking data is available in consumer-grade VR. 
In order to obtain the upper body pose for the arms, we can use the hand controllers as end effectors for an Inverse Kinematics algorithm. This solution is fast to compute and provides a good solution for the user to interact with the environment in VR.

If the animation database does not contain enough variation in velocities and trajectories, the search will constantly return motions that slightly deviate from the target trajectory, which together with the method to avoid positional error explained in Section \ref{sec:position_accuracy} may cause considerable foot sliding. While one possible solution would be to add even more motions to the animation database, this is not always possible due to computation or memory requirements. Therefore, we propose to apply foot lock to improve the final result. The following sections explain how we apply IK and foot lock methods in detail.

\subsection{Upper Body Inverse Kinematics}

Current VR applications use inverse kinematics to adjust the upper-body poses to the hand-held controllers. 
When having only three trackers, using IK alone for full-body (or upper-body) avatars may not produce satisfying results because too much data is missing for several body parts. Consider the case where the body and the head directions are orthogonal (i.e. head looking to the side); this would require us to correctly orient the body so that the IK can find a natural solution to reach the end effectors. Simply using the HMD's forward direction to orient the body will lead to incorrect shoulder locations and the IK will not find a pleasant arm pose. In our work we combine the body orientation predictor and the result given by Motion Matching, which contains the correct body orientation, with IK solvers for the final adjustment of the arms to have the virtual hands following the controllers. \review{Nonetheless, suppose the position accuracy is set to low (e.g., 30\;cm), and in an instant of maximum body position error, the user stretches the arms in the opposite direction. In that case, the IK will fail to reach the target, and the arms will be stretched.} 

\subsection{Foot lock}
As discussed in Section \ref{sec:position_accuracy}, to avoid the virtual avatar deviating too much from the user's position, we constrain the position of the virtual character to follow the position of the user when the error is above a configurable threshold. \review{Thus, the character may be dragged towards the user's position, which causes foot sliding as it is translating the root of the skeleton. We apply foot lock to minimize this issue: the feet will remain locked until a maximum distance is reached or Motion Matching changes the pose.} 

When a pose database is created (Section \ref{sec:pose_database}), foot contacts $\mathbf{y^c}$ are calculated based on the toe’s joint velocity. We store a Boolean value for each foot set to $true$ when the velocity is close to zero. Then, it is used to decide whether each foot should be locked at runtime. IK is used to lock the foot, but to avoid sudden changes, it only applies adjustments to the pose returned by Motion Matching, instead of finding an IK solution from scratch. Finally, if the returned pose and the lock position are too far apart, we unlock the foot to avoid unnatural poses.

\section{Results}
We have implemented the proposed method for animating avatars using the game engine Unity 2021.2.13f1 and PyTorch 1.11.0. We tested it on Oculus Quest 1 and 2, and HTC Vive Pro driven by a PC equipped with an Intel Core i7-8700k CPU, 32GB of RAM and a NVIDIA GeForce GTX 1070 GPU. We used FinalIK from RootMotion \cite{FinalIK} as IK solver to animate the upper body pose. In this section, we compare our method to current solutions for consumer-grade VR used in video games and applications. We analyze the effect of changing the size of the animation database on the position accuracy, and also show how bounding the positional error affects the animation quality. Finally, we analyze independently the body orientation prediction. The readers are referred to the supplementary video for comprehensive comparisons. 

\subsection{Comparisons}
In this section we highlight the main advantages of our method when compared against standard solutions that can be found in current VR applications. We focus on four categories of motions that are common movements in VR. 

\begin{figure}[hbt]
  \centering
  \includegraphics[width=0.7\linewidth]{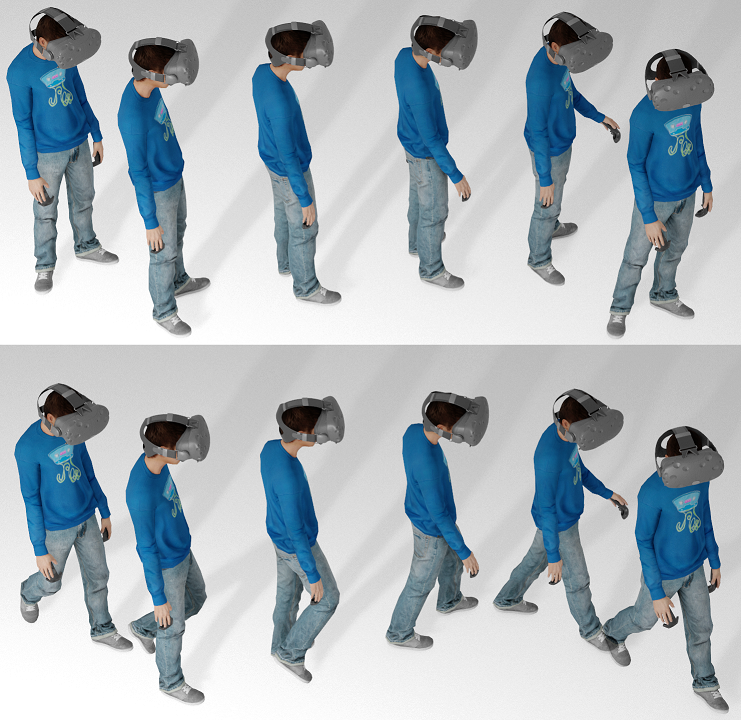}
  \caption{\label{fig:games_vs_ours} A typical approach in VR games consists of simply dragging a static avatar (top). Our solution results in natural walking animations (bottom).}
\end{figure}

\paragraph*{Walking}
When the user is physically walking, most VR applications rendering full-body avatars either drag the character, procedurally generate feet position and apply IK for the leg animation, or apply a fixed animation. These solutions introduce highly noticeable foot sliding and look artificial. In contrast, our solution combining Motion Matching with orientation prediction produces natural-looking walking animations with smooth transitions between different velocities and orientations, thus better adjusting the avatar’s movement to the user (see Figure \ref{fig:games_vs_ours}).

\begin{figure}[hbt]
  \centering
  \includegraphics[width=0.9\linewidth]{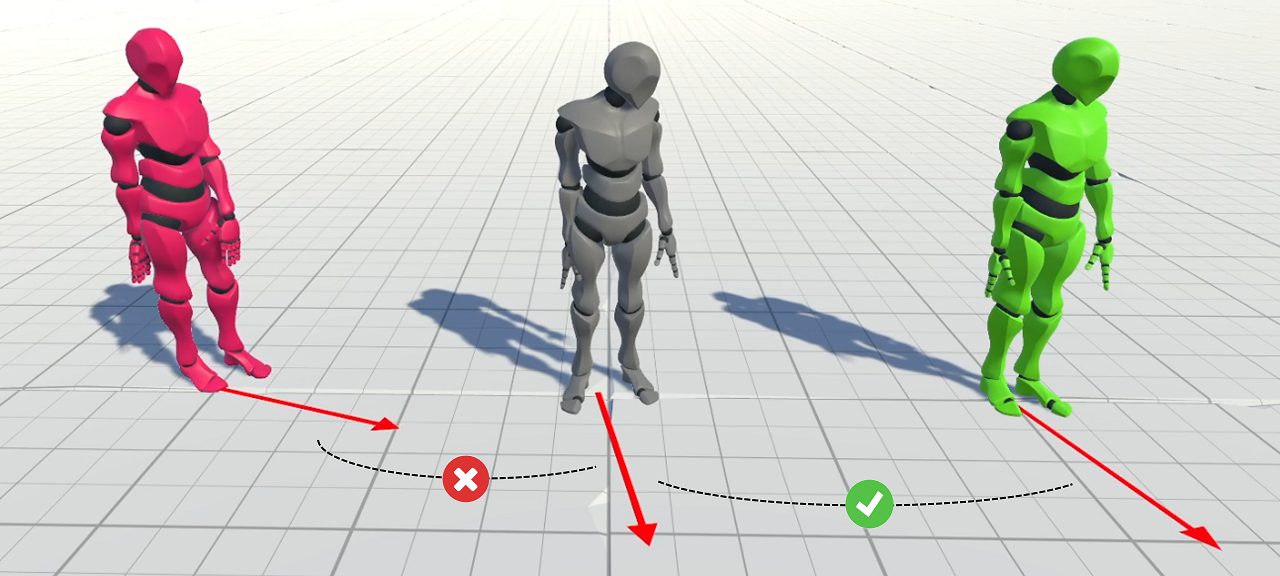}
  \caption{\label{fig:hmdforward_vs_network} In the center the ground truth using Xsens with the red vector indicating the correct body orientation. On the left incorrect torso orientation when using the HMD forward vector as the body orientation. And on the right our neural network prediction.}
\end{figure}

\paragraph*{Body and head orientations}
When using natural walk to navigate in immersive VR, it is important to keep the head and torso orientation decoupled, so that the user is free to move in any direction while rotating the HMD to look around. HMD velocity should also be distinguishable from torso movement, so that the user can take steps in any direction: from walking forward to side stepping. 
Unfortunately, given the lack of torso tracking in consumer grade VR devices, most applications use the HMD’s forward direction to orient the avatar’s body, or simple rotate the avatar’s body when the angle between the HMD’s direction and the current body direction is above a certain threshold. This keeps the avatar torso from being correctly aligned with the user, and often results in wrong motions when applying procedural animation. Our neural network predicts the user’s body orientation from the trackers’ data so that the body can be oriented correctly when combined with Motion Matching. 
Figure \ref{fig:hmdforward_vs_network} compares the resulting avatar orientation with just the HMD (left), a MoCap-based ground truth (center) and our neural network prediction (right). Better orientations in turn lead to smoother animations when using Motion Matching.

\paragraph*{Non-upright motion}
Another important aspect when animating full-body avatars is their behavior when users bend their legs. The applications that allow the avatar to crouch typically require the user to press a button or use the HMD’s height information to bend the legs procedurally. These approaches usually result in incorrect pose matching and unnatural poses. With our approach, non-upright motion (e.g., tip-toes or crouching) is achieved by changing the animation database, which can be switched at run-time based on a given parameter or condition. Additional databases can be included for other type of motions such as dancing, or walking with different gaits. This provides natural non-upright motion with almost no manual setup. Compared to procedural approaches, it preserves the realism of motions as Motion Matching does not synthesize new animations (see Figure \ref{fig:leg_bending}).

\paragraph*{In-place rotations}
In-place rotations are another common movement in VR that the user performs often while looking around, and they typically require small steps for changing the body orientation.
Most applications handle this user movement by rotating the avatar in place keeping a static pose or by applying a slow walk forward animation, but both cases result in noticeable foot sliding. 
Motion Matching naturally handles changes in direction that require small steps, since it is part of the trajectory features in the query vector, thus allowing us to replicate such behavior. 

\subsection{Animation database effect on the position accuracy} 

In Motion Matching, the positional accuracy of the search results is limited by the discrete number of animations in the database. In Section \ref{sec:position_accuracy} we explained the challenge of enforcing the position of the avatar to a specific location when animating self-avatars with Motion Matching, which is aggravated in VR due to the highly unpredictable nature of the user trajectory. Consequently, to keep the avatar at a reasonable distance from the user, we bound the positional error between the user and the avatar and clamp the avatar position if necessary (thus, introducing some foot sliding).

\begin{figure*}[h]
  \centering
  \includegraphics[width=1\linewidth]{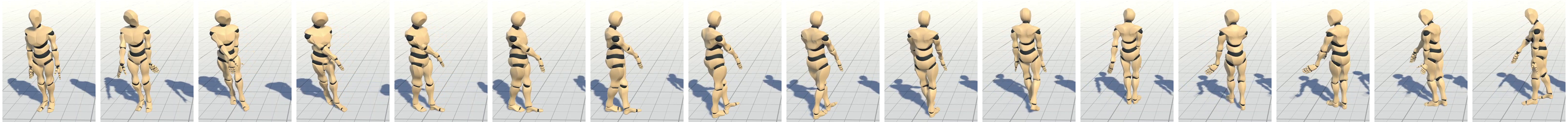}
  \includegraphics[width=1\linewidth]{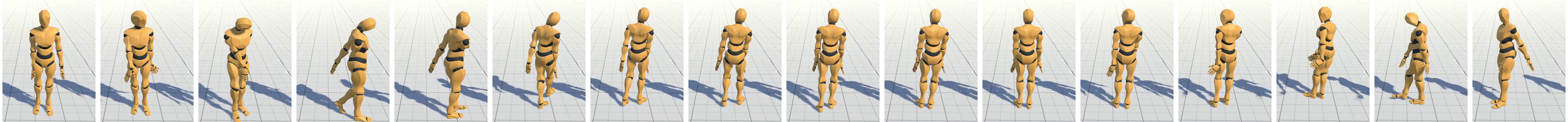}
  \includegraphics[width=1\linewidth]{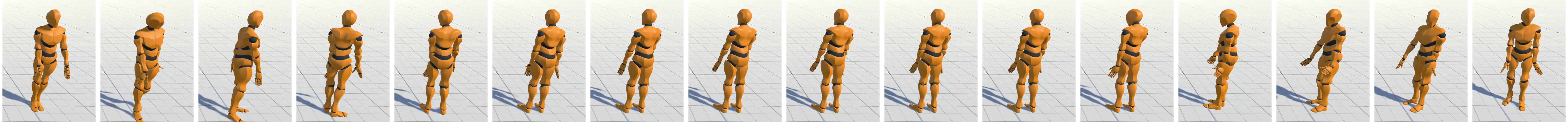}
  \caption{\label{fig:animationdb_comparison} Three avatars animated with the same input (user doing a turn-in-place), but using an increasingly large portion of the animation database. From top to bottom: 10\%, 25\% and 100\% of the poses in the database. Smaller databases struggle to match the user's motion, thus reducing the final quality and increasing the time and space needed to complete the turn. }
\end{figure*}

Ideally, if the animation database could represent all possible user movements in VR, Motion Matching could always perfectly follow the user, and there would be no positional error. Therefore, there is a strong dependency between the positional accuracy of Motion Matching and the size and variety of the database. Figure~\ref{fig:animationdb_comparison} shows the effect of applying different sizes of animation databases to the same user input to show the importance having a good database on the final quality of the movements and the positional accuracy.

\begin{figure}[hb]
  \centering
  \includegraphics[width=0.8\linewidth]{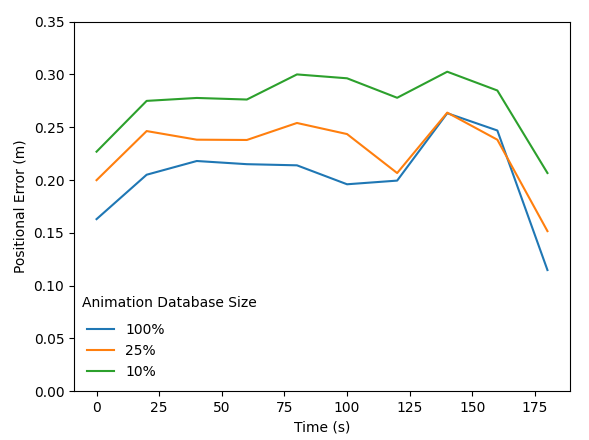}
  \caption{\label{fig:animationdb_positionalerror} Positional error (distance between the user and the avatar) for different animation database sizes using the same user input. 100\% is our complete animation database, while the others contain 25\% and 10\% of the poses in the complete database. The error is averaged every 20 seconds. Large databases represent a wider range of motions and better match the user's trajectories, thus minimizing the positional error.}
\end{figure}

We tested the system with our complete animation database and two reduced versions with 25\% and 10\% of the poses. The position accuracy $\alpha$ was set to 30\,cm. We performed different types of locomotion including run, walk, turn in place, walk in circles, and step sideways, while running Motion Matching for each database. For every frame, the positional error was computed as the distance between the target position (user) and the position of the virtual root joint (avatar). As shown in Figure \ref{fig:animationdb_positionalerror}, the positional error is reduced as the size of the database increases. Some movements had the same error for the complete database and the 25\% version due to the movement being well represented in both databases. The mean positional error for the complete database was 19\,cm, while for the 25\% and 10\% it was 22\,cm and 27\,cm, respectively.

\subsection{Position accuracy effect on the animation quality}

The position accuracy parameter $\alpha$ defined in Section \ref{sec:position_accuracy} is an upper bound of the positional error between the user and the avatar. On the one hand, this parameter should be as low as possible to maintain a good match between the user and the avatar positions. On the other hand, Motion Matching may need some flexibility to find a suitable trajectory to reach the target. The search may not find an exact match to the target trajectory at all times. It may sometimes deviate from the target, causing a positional error, but eventually, it will correct the deviation by searching for new trajectories towards the target. If a maximum positional error is enforced before Motion Matching corrects the trajectory, the animation quality could be reduced due to the limited number of poses used.

\begin{figure}[h]
  \centering
  \includegraphics[width=0.75\linewidth]{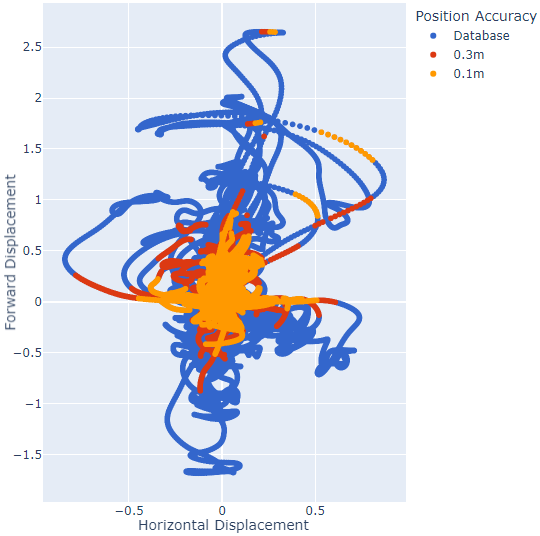}
  \caption{\label{fig:coverage} Poses used by two avatars with $\alpha=0.3$\,m and $\alpha=0.1$\,m were recorded for 5 minutes. The plot shows one of the 2D position features (0.33\;s in the future) from a top-down view. Larger position accuracy allows Motion Matching to use more variety of poses.}
\end{figure}

We set up two avatars with $\alpha=0.3$\;m and $\alpha=0.1$\;m, and controlled them simultaneously for around 5 minutes. We recorded the indices to the pose and feature databases used for each avatar. Figure \ref{fig:coverage} shows one of the 2D position features (0.33 seconds in the future) for all poses in the database and for those poses used for the avatars. The avatar with $\alpha=0.3$\;m has more freedom of movement and can use a larger number of poses, thus, enhancing the final animation quality. 
In total, the avatar with $\alpha=0.3$\;m used 6,932 different poses while the other used 5,715 different poses. 

\review{Consequently, a large $\alpha$ is recommended for other users' avatars to improve the animation quality and reduce visual artifacts such as foot sliding. However, the quality of certain types of motion (e.g., reaching an object) may be reduced because of positional misalignment: upper body IK will frequently fail to reach the target. Visual artifacts on the legs are less noticeable when animating self-avatars, as users usually do not look at their legs \cite{lee2020}. Thus, we suggest using a small $\alpha$ to avoid problems with the IK and the location of the virtual body for self-avatars. }

\subsection{Body orientation prediction}

In this section, we compare our neural network and the HMD’s forward direction for predicting body orientation. We also compare the accuracy of the neural network depending on the parameter $r$ defined in Section \ref{sec:network_training}. Xsens was used to capture the motion of a user playing 
a video game for Oculus Quest for 15 minutes. The game required the user to change the direction of the body and the head frequently. We measured the angle error between the ground truth body direction captured with Xsens, which is the projection of the hip joint forward direction onto the ground, and the different body orientation predictors. 

\begin{figure}[htb]
  \centering
  \includegraphics[width=0.8\linewidth]{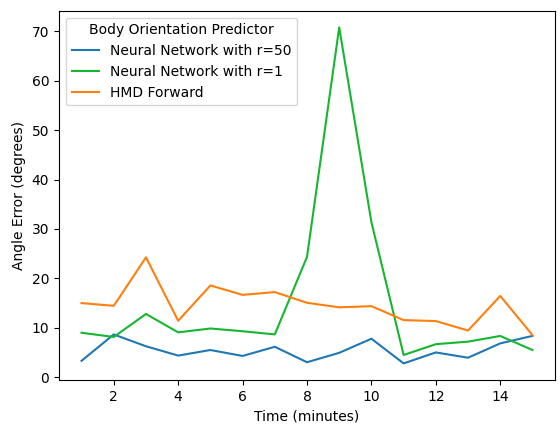}
  \caption{\label{fig:body_direction_predictors} Average angle error per minute for different body predictors while playing a video game for 15 minutes. HMD's forward direction had a larger error than our neural network for predicting the body orientation. The neural network trained with $r=1$ learns to imitate the previous orientation instead of using the trackers’ data resulting in a larger error than training with $r=50$.}
\end{figure}

Figure \ref{fig:body_direction_predictors} shows the average angle error per minute for the different body predictors. Directly using the HMD’s forward direction as body direction had a mean angle error of 14.5º and a standard deviation of 18.9º while using our neural network trained with $r=50$ had a mean angle error of 5.4º and standard deviation of 7.7º. When $r=1$, the neural network learns to imitate the previous orientation, which is one of the network's inputs, because it always comes from the ground truth data during training. At runtime, the previously predicted orientation is given as an input to the network, therefore, it may not be reliable. Instead, we want the network to learn to use the tracker’s information to predict the new orientation while still having access to the previous one to induce continuity in the changes. When $r=50$, the neural network is trained to predict 50 consecutive frames and only compares the last one with the ground truth data, which makes the network use the tracker's information to predict the future orientations as simply copying the previous predicted orientation is not enough. The mean angle error when $r=1$ is 15.0º, and the standard deviation is 25.7º. The result is worse than when using $r=50$ because, as shown in Figure \ref{fig:body_direction_predictors}, around minute 7, the angle error is very large, and since the neural network is imitating previously predicted orientations, it cannot quickly recover from the failure.

\section{Discussion and limitations} 
\label{sec:discussion}


\paragraph*{Why not animate the upper body with Motion Matching}
One of the strengths of Motion Matching is the final high-quality animations that it can provide directly from the motion capture database. However, new animations are not synthesized from the existing data. The search finds the best sequence of poses that approximate our target trajectory, but since it is unfeasible to have a database with all possible movements (and with different speeds, styles...) 
it is common to add procedural touch-ups such as slightly rotating the virtual character towards the desired direction. Responsiveness and accurate positioning of the arms are crucial when it comes to VR. Therefore we would need the position (and orientation) of the controllers (or similar features) to match the upper body pose. However, users in VR are free to do any movement with their arms, and to represent the 6 degrees of freedom of each controller accurately, the number of different trajectories and poses needed in the animation database would make it unmanageable due to its size.

Even if a large animation database was available, we found through exhaustive testing that the high dimensionality of the feature vector introduces a challenge to the search when prioritizing features and as a result it does not return continuous movements. We believe more research is necessary to deal with these issues. For instance, neural networks can be used to effectively represent massive databases as in \cite{holden2020}, and eventually learn the motion manifold to overcome the limitation of having a discrete set of poses in the animation database.



\paragraph*{Quality dependency on the database}
The main component of our method is Motion Matching, which provides high-quality locomotion animations, and could be directly incorporated in most existing VR applications using full-body avatars. To cover most of the recurrent motions in VR, the creation of the database is critical: it should contain a good variety of walking speeds, a high number of in-place turns, and sideways walking. The behavior of the system can be easily modified by changing the animation database, for instance, different walking gaits (sad, happy, hurt, etc.) could be represented just by changing the database, or we could use a dancing database to make the lower-body of the avatar dance according to the movement of the HMD and the body direction (see accompanying video).

\paragraph*{Upper body animation}
This paper does not focus on the upper body IK because it is already incorporated in any application using full-body avatars. However, by combining orientation prediction with Motion Matching before applying upper body IK, we get two benefits. First, the avatar will always have a correct pose for the torso, thus resulting in a better arm position. Second, if we detect a controller failure, we could disable IK for that arm and let Motion Matching return a plausible pose for the arm.

\paragraph*{Foot lock}
Motion matching suffers from the foot sliding problem when handling hard constraints with procedural touch-ups (e.g., enforcing an exact position for the character). This issue arises from the discrete nature of the animation database. The problem is typically alleviated with IK to lock the foot to the floor. However this problem becomes more noticeable in VR due to the highly unpredictable nature of the user trajectory in real time.

In the case of animating characters to follow a path or a joystick input, the trajectory used for Motion Matching is the exact trajectory that the character has to follow. However in VR, predicting such trajectory depends on the real time movement of the user which can rapidly change, and on the velocity of the HMD, which is irregular and may suffer from latency  (e.g., remote collaborators). This difficulty to create accurate trajectories introduces more noticeable foot sliding problems when enforcing hard constraints. Moreover, the problem is further aggravated by the larger amount of locomotion movements that can be performed in VR compared to animating a character in a video game. %
Still, we observed that foot lock could successfully minimize foot sliding, and we believe that it could be disabled as the animation database grows, allowing the system to find better matches to the target trajectory, or when the maximum positional error allowed is large enough.

\section{Conclusions and future work}

With the increasing interest in using avatars for the user representation in VR applications, such as video games, collaborative meetings or the Metaverse, there has been a growing amount of research in areas such as facial animation or 3D avatar reconstruction from images. However, body pose animation is still relying on traditional animation methods. We believe our approach can help improve current VR applications by providing higher-quality animations for avatars. 

In this work, we have presented a data-driven method combining body orientation prediction and Motion Matching for animating full body avatars in mainstream VR devices (i.e., one HMD and two controllers). Our system can improve existing VR applications using full-body avatars since it provides good animations that correctly follow the user movements without requiring additional tracking devices. To have a good pose alignment between the avatar and the user body, we present a neural network for predicting the body orientation and use it as input for Motion Matching. Overall, we hope this work will help in the development of high-quality data-driven animations in the field of VR that can greatly improve embodiment and facilitate collaborative work in VR.  


For further research, it would be interesting to incorporate the upper body animations into the Motion Matching system by reducing the dimensionality of the feature vector or combining multiple searches for different body parts. The use of deep learning can be another interesting direction of research to increase the expressiveness of the method and avoid the limitations of using a discrete number of poses in the animation database.

\paragraph*{Code and data}
The complete source code, trained model, animation databases, and supplementary material used in this paper can be found at
\href{https://upc-virvig.github.io/MMVR}{https://upc-virvig.github.io/MMVR}.

\section{Acknowledgements}
This project has received funding from the European Union’s Horizon 2020 research and innovation programme under the Marie Skłodowska-Curie grant agreement No. 860768 (CLIPE project) and the Spanish Ministry of Science and Innovation (PID2021-122136OB-C21).


\printbibliography


\end{document}